\pdfoutput=1
\documentclass[11pt]{article}

\usepackage[T1]{fontenc}
\usepackage[utf8]{inputenc}
\usepackage{booktabs}
\usepackage{amssymb}
\usepackage{mathptmx} % Times text and math; universally available (incl. arXiv)
\usepackage[margin=1in]{geometry}
\usepackage{graphicx}
\usepackage{microtype}
\usepackage[font=small,labelfont=bf]{caption}
\usepackage[colorlinks=true,linkcolor=blue,citecolor=blue,urlcolor=blue]{hyperref}
\usepackage[numbers]{natbib}

\hypersetup{pdftitle={Hash Table Design for RDMA: Challenges and Opportunities},pdfauthor={Shuchen She}}

\title{\textbf{Hash Table Design for RDMA:\\Challenges and Opportunities}}

\author{
Shuchen She, Hancheng Wang, Haipeng Dai, Ruofei Ju, \\ Bohao Liu, Jingye Su, Yuxuan Xie, Yutian Zhang\\[2pt]
\textit{Nanjing University}\\
\texttt{sheshuchen711@gmail.com}
}

\date{} % arXiv adds its own date stamp; leave empty or set a date here.

\begin{document}

\maketitle

\begin{abstract}
Hash tables complete the insertion, lookup, and deletion of a single key in constant time on average, and they are widely used in databases, key-value stores, and network systems. In the Internet of Things (IoT), the number of devices and the volume of sensed data keep growing, so the hash tables that store or index the data consume more and more memory. When a single server runs out of memory, the system can place part of the data in the memory of other nodes. One-sided operations in Remote Direct Memory Access (RDMA) let one machine read and write the memory of another machine directly, with low latency and high bandwidth, and are therefore widely used to build disaggregated memory systems. Deploying a hash table on RDMA-based remote memory exploits the memory of other nodes and thus relieves the capacity limit of a single server. However, this deployment raises three problems. First, one logical hash-table access may translate into one or more remote network accesses, and collision handling and probing further increase the number of RDMA requests. Second, because the remote CPU is bypassed, traditional concurrency control that relies on remote threads no longer applies directly. Third, the limited resources of RDMA network interface cards, such as queues, caches, memory registration, and atomic operations, impose new constraints on hash table structures. This paper focuses on hash table design for RDMA. We review existing work, distill the key challenges, and discuss promising optimization directions and coping strategies, aiming to provide a reference for designing remote hash tables in IoT big-data scenarios.
\end{abstract}


\section{Introduction}

A hash table maps each key to a storage location through a hash function. It completes the insertion, lookup, and deletion of a single key in constant time on average. Thanks to this fast key-value access and point-query capability, hash tables and other hash-based techniques are widely used in databases, key-value stores, network systems, and machine learning~\cite{carlson2013redis,gomez2018water,li2024istack,ratnasamy2006multicast,yu2022ibm,yu2021bidi,wang2019multifocus,Dayan2021Chucky,Dayan2017monkey,zhang2018surf,Redis2026Redis,alluxio2025alluxio,scyllaDB2025scyllaDB,cassandra2025cassandra,partitioned2024partitioned,tikv2025tikv,google2023leveldb,facebook2021rocksdb}. First, distributed key-value stores often use hash tables for fast key lookups, for example to manage Internet of Things (IoT) data such as device states and sensor readings~\cite{carlson2013redis,gomez2018water}. Second, in packet forwarding and flow-table management, hash tables serve exact-match lookups that quickly locate the matching flow entry or forwarding rule~\cite{li2024istack,ratnasamy2006multicast}. Third, for data mining and complex event recognition over IoT data streams, structures such as hash tables, filters, and bitmap indexes record the occurrences of data items, narrow the set of candidate events, and reduce unnecessary scans~\cite{dai2016persistent,liu2026bitmap,yu2021cart,yu2019ocs,yang2023basictad}. Because hash tables play a central role in so many systems, researchers have proposed a variety of designs that optimize lookup throughput, insertion efficiency, space utilization, concurrent access, and cache friendliness~\cite{litwin1980linear,herlihy2008hopscotch,zuo2018level,lu2020dash,nam2019cceh}.

As IoT deployments scale up, the amount of data that systems must store and index grows rapidly~\cite{yang2024adapting,zhou2024rts,yang2025mobileviclip}. Sensor data streams are produced continuously and keep accumulating, and the associated metadata, key-value mappings, and query indexes expand with them. In such systems, hash tables typically support fast lookup of device identifiers, sensor readings, or metadata, so their capacity requirement grows with both the data volume and the index size. Meanwhile, data-intensive applications already report processing demands at the level of hundreds of terabytes~\cite{gupta2021rambo}. When the data volume exceeds the memory capacity of a single server, the system can adopt a disaggregated memory architecture and place part of the data and index structures in the memory of other nodes, thereby breaking the single-server capacity limit. How to use the memory of other nodes efficiently to extend the capacity of hash tables and related index structures has therefore become a practical problem in IoT data infrastructure.

To access the memory of other nodes efficiently, modern datacenters increasingly adopt Remote Direct Memory Access (RDMA). After memory registration and connection setup, RDMA allows an application to read and write registered remote memory regions directly through the network interface card, bypassing the kernel network stack and the remote CPU on the data path~\cite{dragojevic2014farm,mitchell2013pilaf}. Compared with traditional TCP/IP-based networking, RDMA usually offers lower access latency, higher throughput, and lower CPU usage~\cite{kalia2016guidelines}. On top of RDMA and other high-speed networks, datacenters have further developed the disaggregated memory architecture~\cite{gao2016network,maruf2023memory}. This architecture separates compute resources and memory resources into a compute pool and a memory pool. The compute pool has strong computing power but limited local memory. The memory pool provides abundant memory but weak computing power, and it is generally unsuitable for complex operations~\cite{maruf2023memory}. The compute pool accesses the remote memory in the memory pool through RDMA. Under this architecture, the data region of a hash table and its related index structures can reside in the memory pool and be accessed by the compute pool via RDMA.

However, directly porting a hash table from a single server to RDMA-based remote memory causes significant performance problems. First, random accesses of a single-server hash table mostly hit local memory. On remote memory, the same accesses become remote operations such as RDMA READ, RDMA WRITE, or atomic operations, and each remote operation pays a network round trip. The repeated random probing, collision handling, and resizing inside a hash table therefore become much more expensive. Second, RDMA provides only low-level communication primitives. It offers no hash-table-level semantics for lookup, insertion, deletion, migration, or consistency, so the clients must build the remote data layout, the concurrency control scheme, and the consistency mechanisms by themselves. Third, the hardware resources of the RDMA network interface card (RNIC), including queues, caches, address translation, and atomic operations, are limited. These limits constrain the data layout, the access granularity, and the concurrency control of a hash table.

This paper studies hash table design for RDMA. We review the design ideas of existing remote index structures and distill the key problems they face on remote memory. Our contributions are threefold. First, we organize existing studies around remote access overhead, concurrency and consistency, and dynamic resizing, and we discuss how filter structures can help optimize remote index lookups. To the best of our knowledge, a systematic survey dedicated to hash table design for RDMA remains missing. Second, we identify several open problems that RDMA hash tables have not fully solved: the high round-trip cost of remote accesses, the network amplification caused by collision handling, the complexity of maintaining consistency once the remote CPU is bypassed, and the difficulty of dynamically resizing large index structures. Third, to address these problems, we summarize the corresponding optimization directions and propose design principles, including reducing the number of remote accesses, improving data locality, lowering concurrency synchronization overhead, and supporting efficient resizing, as a reference for future research.

The rest of this paper is organized as follows. Section~\ref{sec:background} introduces the background on RDMA and memory disaggregation, and reviews existing work on hash tables and filters. Section~\ref{sec:challenges} analyzes the main challenges in hash table design for RDMA and the corresponding solutions. Section~\ref{sec:conclusion} concludes the paper and outlines future research directions.

\section{Background and Related Work}
\label{sec:background}

To set the stage for RDMA-based hash table design, this section first introduces the RDMA operation model and the disaggregated memory architecture. It then reviews hash tables for persistent memory, hash tables for RDMA, and related filter studies, highlighting the limitations of existing work that motivate the key challenges discussed in this paper.

\subsection{RDMA Operation Model and Disaggregated Memory Architecture}

\begin{figure}[t]
  \centering
  \includegraphics[width=0.62\linewidth]{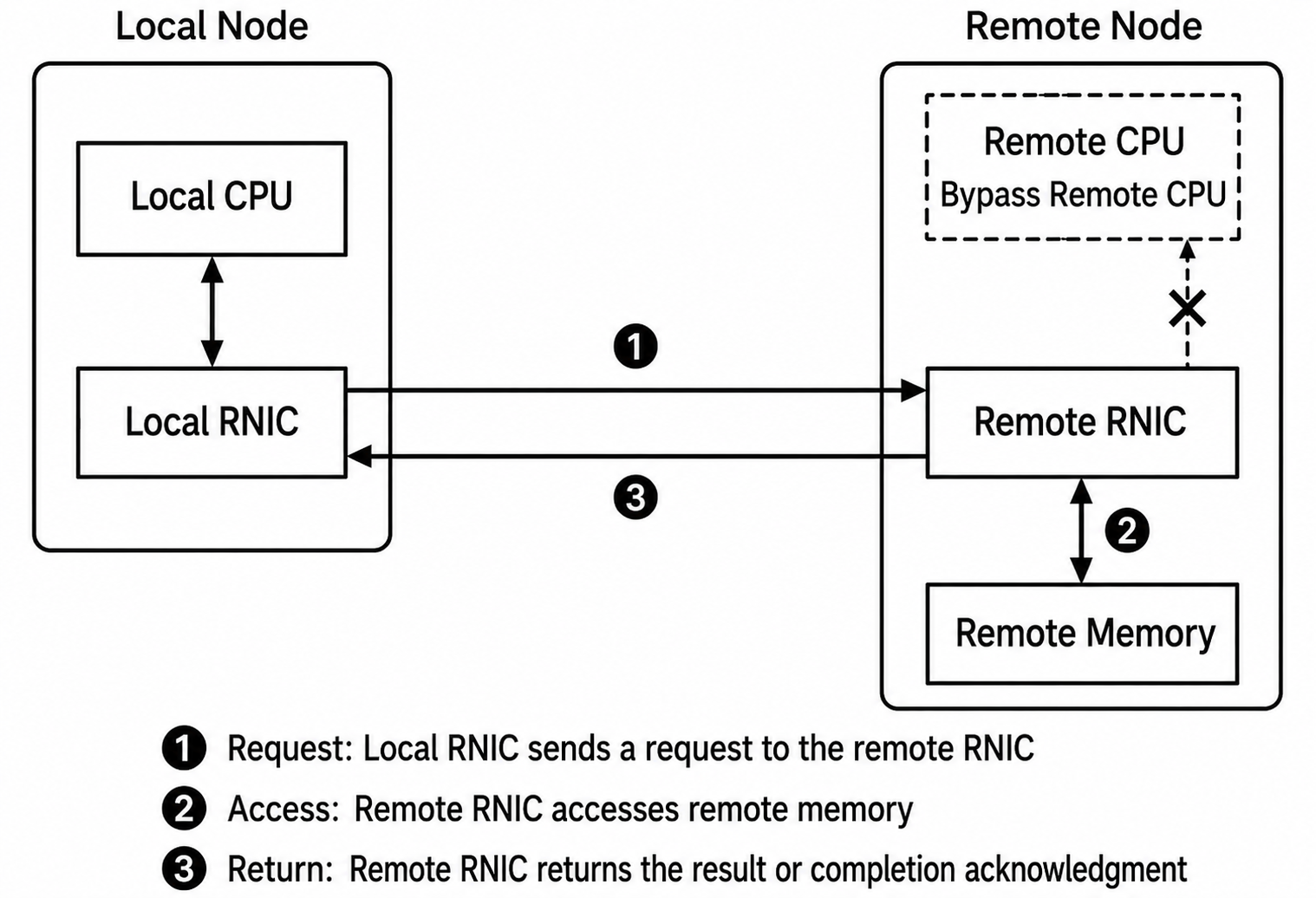}
  \caption{Illustration of RDMA one-sided operations.}
  \label{fig:onesided}
\end{figure}

RDMA data transfers are executed mainly by the RNIC, which supports two common classes of operations~\cite{min2024sephash}. The first class is two-sided operations, namely send and receive. Two-sided operations require the receiver to post receive requests in advance; after a message arrives, the remote application usually still has to handle the completion event and run the corresponding application logic. The second class is one-sided operations, including remote read, remote write, and atomic operations. Among the atomic operations, compare-and-swap (CAS) is commonly used for concurrent insertions into RDMA hash tables. A CAS first checks whether the current value at a remote memory location equals an expected (old) value; if so, it replaces the value with a new one. Take SepHash as an example~\cite{min2024sephash}: each entry in a CurSegment is 8 bytes, and clients use RDMA CAS to compete for an empty entry, which guarantees that no two clients can both succeed in writing the same location. One-sided operations need no active participation of the remote CPU; the initiator accesses the registered remote memory region directly through its local RNIC. As shown in Fig.~\ref{fig:onesided}, the request of a one-sided operation travels from the local RNIC to the remote RNIC, the remote RNIC accesses the remote memory and returns the result or a completion acknowledgment, and the whole data access bypasses the remote CPU.

In an RDMA network, one remote access usually takes microseconds, whereas one local memory access takes nanoseconds; the two differ by a factor of tens to more than a thousand. Reducing the number of remote accesses and the request submission overhead is therefore a primary goal of RDMA data structure design. Because this operation model bypasses the remote CPU and accesses remote memory directly, RDMA has become a key building block of disaggregated memory architectures. Memory disaggregation is a datacenter architecture that deploys compute resources and memory resources separately~\cite{gao2016network}. The system consists of a compute pool and a memory pool, connected by a high-speed network such as RDMA. The compute pool consists of compute nodes with strong computing power; each node has CPUs and some local memory, but this local memory is relatively small and mainly holds caches, runtime state, and temporary data~\cite{zuo2021race}. The memory pool provides a large amount of remotely accessible memory but has weak computing power; it mainly performs lightweight tasks such as network communication, memory management, and metadata maintenance~\cite{liu2026cer}. Separating compute from memory alleviates the resource waste caused by the fixed coupling of CPU and memory in traditional servers, improves resource utilization, and makes the system more elastic under resource expansion and load changes~\cite{zuo2021race,dou2026intelligence}.

Under the disaggregated memory architecture, the compute pool accesses the remote memory in the memory pool through RDMA. The lookup, insertion, and deletion logic of the hash table is executed entirely in the compute pool, which performs all accesses through RDMA. The hash table can then exploit the scalability of remote memory and alleviate the memory-capacity limitation of a single server. To reduce the local overhead of submitting multiple consecutive RDMA operations, RDMA supports doorbell batching. With this mechanism, the application hands all requests to the RNIC at once, which reduces the number of doorbell writes and the local submission overhead. In addition, during RDMA accesses to remote memory, the RNIC must validate and translate the target address according to the address mappings of the registered memory regions. The RNIC caches part of the address translation information internally, which removes most of the translation overhead from each access. When the hash table is large and spans many remote memory pages, the number of address mappings grows accordingly. Once the mappings exceed the capacity of the RNIC's address-translation cache, the RNIC has to fetch the mapping information again, which increases access latency and degrades overall performance.

Beyond RDMA-based memory disaggregation, emerging interconnects such as Compute Express Link (CXL) are also advancing remote and expanded memory. A recent study revisits hash table design for CXL memory and shows that, on a memory medium with higher access latency, different bandwidth characteristics, and coherence overhead, traditional hash tables face new challenges in access granularity, metadata organization, concurrency control, and update mechanisms~\cite{wang2024cxl}. This finding indicates that hash table design cannot rely only on the cost model of local memory; it must be re-optimized for the underlying memory interconnect and the characteristics of remote access.

\subsection{Hash Tables for Persistent Memory}

Before analyzing remote hash tables for RDMA, we briefly review hash table research on persistent memory. Persistent memory is a large-capacity, byte-addressable memory that retains data across power failures, and prior work has examined memory access behaviors and NUMA effects on tiered NVM systems~\cite{yang2020monitoring}. Understanding how hash tables are adapted to its access characteristics helps explain why hash tables must be redesigned for RDMA. Existing studies improve persistent memory hash tables mainly in terms of resizing, concurrency, recovery, and performance stability.

First, researchers have proposed various structures to reduce the data movement caused by resizing. The most straightforward resizing method is full-table rehashing, which remaps all existing items into a new, larger table; this incurs heavy data movement and high tail latency. To address this problem, Level hashing adopts a two-level structure and moves only one level of data during a resize, which reduces the amount of data moved per resize~\cite{zuo2018level}. CCEH borrows the idea of extendible hashing~\cite{nam2019cceh,fagin1979extendible} and manages buckets or segments through a directory, so resizing can be performed locally instead of rebuilding the whole table; this lowers the impact of resizing on normal accesses.

Second, researchers have also improved persistent memory hash tables in terms of concurrent access and crash recovery. Clevel extends Level hashing with a lock-free concurrent design, which reduces the overhead caused by lock contention~\cite{chen2020clevel}. Dash reduces crash recovery time and uses lazy recovery to shorten the time before the hash table becomes available again~\cite{lu2020dash}. Halo identifies the write amplification caused by the mismatch between the write granularity of persistent memory and the actual update granularity~\cite{hu2022halo}. Pea hash adaptively trades off throughput and space overhead~\cite{liu2023pea}. SEPH targets stable throughput and fine-grained resizing~\cite{wang2023seph}. However, these hash tables are built for local persistent memory and cannot be applied directly to RDMA scenarios. We next introduce hash table designs built on RDMA remote memory.

\subsection{Hash Tables for RDMA}

Persistent memory hash tables mainly tackle the costs of writes, recovery, and resizing on a locally addressable medium. RDMA instead places hash table accesses on the network path, so each bucket access turns into one or more remote network round trips. A hash table for RDMA must therefore consider not only its own space utilization and concurrency efficiency, but also the number of remote accesses, the access granularity, the semantics of one-sided operations, and the limited computing power of the memory pool. Some RDMA hash tables and key-value systems focus on the remote access efficiency of the lookup path: FaRM, Pilaf, and DrTM organize their tables with hopscotch hashing, cuckoo hashing, and cluster chaining, respectively~\cite{dragojevic2014farm,mitchell2013pilaf,wei2015drtm}. Among them, Pilaf mainly uses one-sided RDMA READs to accelerate lookups and reduce remote CPU involvement during lookups, but writes, deletions, and updates still rely on the server CPU or more complex remote coordination. DrTM supports reading and writing remote key-value pairs with one-sided RDMA operations, but its design depends on a transaction system and specific concurrency control mechanisms. In other words, these systems are not designed for a disaggregated memory architecture whose memory pool has almost no computing power.

On disaggregated memory, the memory pool only provides remote memory; its weak computing power cannot carry the complex logic of hash table updates, collision handling, and concurrency control. The insertion, deletion, and update logic of a traditional hash structure must therefore move entirely to the clients, and remote modifications must rely on one-sided RDMA operations alone. As a consequence, collision handling, item migration, and consistency maintenance may generate many remote round trips. For example, cuckoo hashing may repeatedly migrate items under high load factors, and each migration may involve remote reads, remote writes, and concurrency checks. In cluster chaining or chained hashing, the location of the next bucket on a collision chain usually depends on the result of the previous read, so the reads can hardly be batched in advance.

RACE is an early RDMA hash table designed specifically for disaggregated memory. Its goal is to complete lookup, insertion, update, and resizing using only one-sided RDMA operations, without relying on any computing resources in the memory pool~\cite{zuo2021race}. RACE adopts extendible hashing and organizes the table as one directory plus multiple subtables. Each key maps to two candidate main buckets in a subtable, and two adjacent main buckets share one overflow bucket that holds the items the main buckets cannot accommodate. A main bucket and its shared overflow bucket together form a combined bucket, so a lookup can examine the candidate slots in the main bucket and the shared overflow bucket at once. RACE further uses doorbell batching to submit the related RDMA READs back to back, which reduces the local submission overhead and lets a lookup usually finish within a short remote access path. To cut the remote data transfer during lookups, RACE stores in each bucket a fingerprint derived from the hash value of each key. A client first reads the bucket and compares the fingerprints. If a fingerprint does not match, the client rules out that key directly; only when the fingerprint matches does the client read the full key-value data for an exact comparison. This avoids a large number of unnecessary remote reads.

For concurrency, RACE adopts lock-free remote concurrency control, so ordinary lookups and updates run concurrently without acquiring remote locks. Because reads and writes may interleave, a client may observe an intermediate state of an ongoing update; RACE therefore embeds checksums in the key-value data so that the client itself can verify whether a read result is complete and consistent. For resizing, RACE resizes locally: it extends only the subtable that overflows instead of rebuilding the whole table. To avoid extra round trips for directory accesses, RACE caches the directory at the clients. To handle directory caches that become stale after a resize, RACE records metadata in each bucket header for validating the directory mapping. After reading a bucket, the client uses this metadata to check whether the bucket still belongs to the target subtable; if the cached directory is stale, the client refreshes its local cache and locates the bucket again.

RACE may still suffer from resize-induced data movement and remote concurrency control overhead under write-intensive or resize-heavy workloads. To address this problem, SepHash proposes a write-optimized hash table for disaggregated memory~\cite{min2024sephash}. SepHash observes a bandwidth--latency trade-off in the RDMA access granularity. Small accesses transfer little data per request but inflate the number of requests and the protocol overhead, thereby lowering bandwidth utilization. Large accesses improve bandwidth utilization but increase the data volume and the latency of each access. Based on this observation, SepHash designs a two-level separate segment structure that migrates entries in batches during a resize instead of one at a time, which reduces the data movement and write amplification during resizing and saves bandwidth.

For concurrent writes, SepHash adopts an append-based strategy. An insertion does not overwrite the old entry in place; it appends the new version to an empty entry in the currently writable segment. Multiple versions of the same key are laid out in write order, and newer versions occupy later visible positions. This avoids complex in-place updates and remote lock contention, and it lowers the synchronization overhead of concurrent insertions. On the lookup side, a client identifies the valid result using the state, version, or depth information in the entries, which reduces the re-reads and extra round trips caused by concurrent writes. For lookup optimization, SepHash also uses fingerprints, and it combines filters with client-side caching to cut unnecessary remote accesses, so it keeps good lookup performance while optimizing writes.

More recently, several studies further optimize remote index structures from the perspective of key-value systems on disaggregated memory. Outback designs a communication-efficient index for key-value stores on disaggregated memory, focusing on reducing the remote communication during index accesses~\cite{liu2024outback}. FUSEE takes the perspective of a complete key-value store and studies how to organize the index, the caches, and the data access path when memory is fully disaggregated~\cite{shen2023fusee}. These studies do not always propose a new hash table structure, but they further show that on RDMA, a hash table must be co-designed with the system's access paths, caching mechanisms, and remote communication costs.

Besides hash tables, tree-based indexes also exist on disaggregated memory; one example is Sherman, a write-optimized distributed B$^{+}$-tree~\cite{wang2022sherman}. Beyond conventional tree structures, disaggregated-memory systems have also explored learned and radix-tree indexes. ROLEX uses learned models in an RDMA-oriented key-value store, whereas SMART adapts an adaptive radix tree to the disaggregated-memory setting~\cite{li2023rolex,luo2023smart}. These designs show that reducing remote accesses and controlling metadata and update overhead are common concerns across hash, tree, hybrid, and learned indexes. Client-side caching and hybrid index organizations provide two additional ways to reduce the cost of remote indexing. XStore uses a learned cache at the client side to reduce repeated remote index traversals~\cite{wei2020xstore}. Han et al. combine a hash table for efficient point lookups with an ordered index for range queries in an RDMA-based key-value store~\cite{han2023hybrid}. DEX further studies scalable range indexing on disaggregated memory and uses lightweight caching and selective computation offloading to reduce remote accesses~\cite{lu2024dex}. Together, these studies indicate that remote index performance depends not only on the underlying data structure, but also on how metadata, caches, and computation are distributed between compute nodes and memory nodes. Tree-based indexes support ordered access and range queries, whereas hash tables fit exact-match point queries. In the broader indexing literature, multi-dimensional indexes include the R-tree, the KD-tree, the Quadtree, and newer structures designed for modern hardware~\cite{li2024survey}. CHIME further proposes a hybrid index for disaggregated memory that combines the strengths of different index structures to improve cache efficiency and remote access performance~\cite{luo2024chime}.

\begin{figure}[t]
  \centering
  \includegraphics[width=0.95\linewidth]{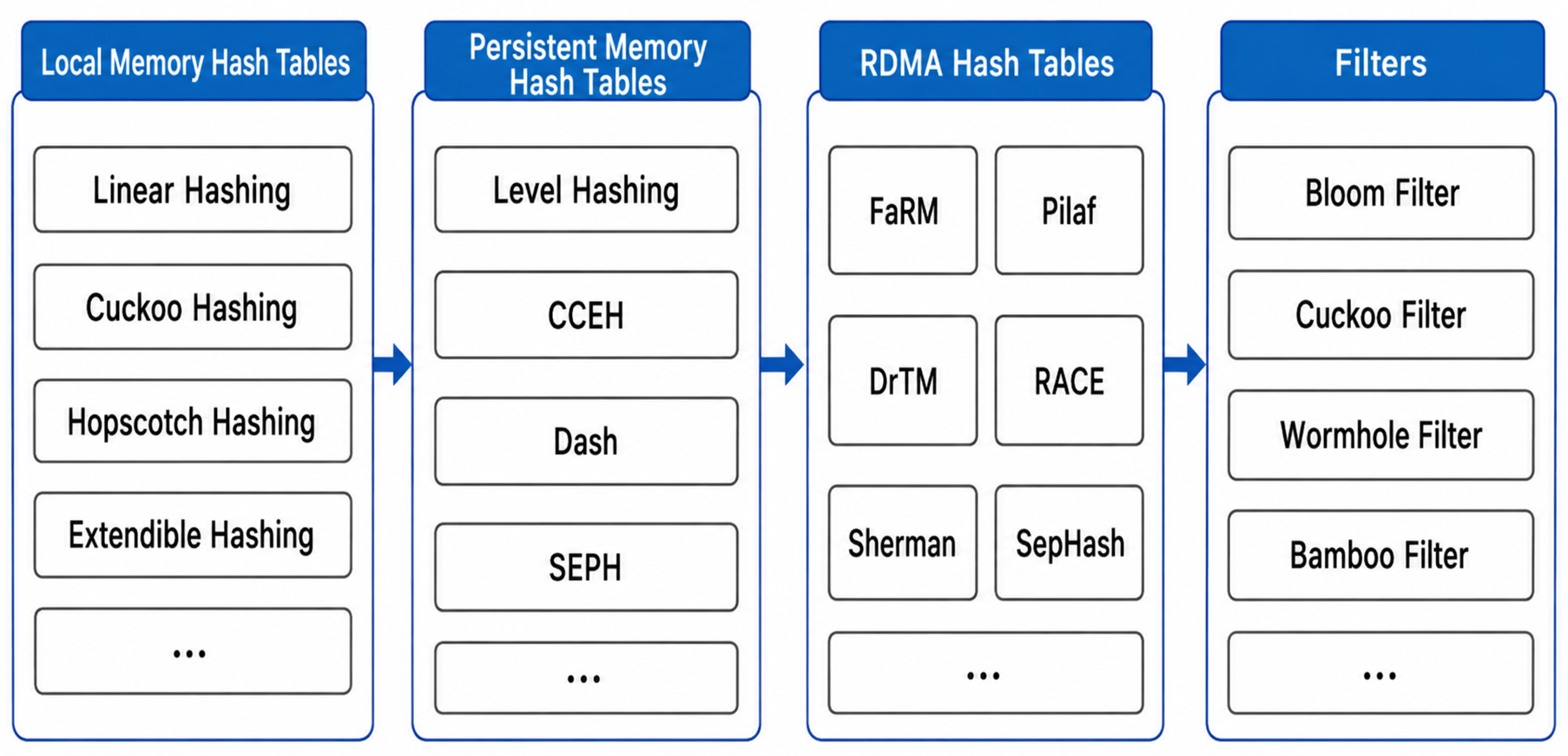}
  \caption{Illustration of the evolution of index structures across different hardware environments.}
  \label{fig:evolution}
\end{figure}

Fig.~\ref{fig:evolution} outlines how typical index structures have evolved from hash tables in local memory, to hash tables on persistent memory, and then to hash tables and filters for RDMA. As the underlying storage medium and access path change, the optimization goals shift accordingly: from purely reducing algorithmic complexity, to reducing the number of remote accesses, controlling the access granularity, lowering the data movement cost, and adapting to hardware resource limits.

\subsection{Filters}

Besides optimizing the hash table itself, reducing unnecessary accesses to the remote hash table is another important optimization for RDMA. Like hash tables, filters are hash-based data structures, but a filter only answers approximate membership queries and stores no complete key-value data. In practice, a filter usually serves as a pre-check before hash table accesses. When the filter reports that a key does not exist, the client skips the remote hash table entirely; when the filter reports that the key may exist, the client then queries the hash table for an exact answer. Filters are therefore closely related to the lookup optimization of remote hash tables. Common filters include the Bloom filter and the cuckoo filter~\cite{bloom1970,fan2014cuckoo}. A filter keeps no complete raw data; it keeps only bits or fingerprints computed from the keys, so its space cost is far smaller than storing the full key set~\cite{fan2000summary}. The price of this space efficiency is false positives: for a key that does not exist, the filter may still report that it exists. In many systems, however, a false positive only causes one extra lookup and never affects the correctness of the result. Filters are therefore commonly used as a pre-check before the actual lookup to cut unnecessary accesses. In IoT, filters are often used to reduce storage and communication costs. For example, in wireless sensor networks, filters assist authentication, membership checking, and duplicate detection; in vehicular networks, filters support content-cache checks to reduce unnecessary cache queries~\cite{mbarek2018bfan,dua2019bloom}.

Some studies build filters specifically for persistent memory. Wormhole filters target the access overhead of filters on persistent memory and reorganize the hash and fingerprint information to reduce that overhead~\cite{wang2024wormhole,wang2025parallel}. Beyond optimizing a specific filter structure, other work improves the accuracy of approximate membership queries through encoding. The variable-length encoding framework assigns codes of different lengths to different items, which increases the expressiveness of a filter and lowers its false positive rate~\cite{dai2023vlef}. These methods show that filter performance depends not only on the bucket layout and the hash access pattern, but also on how the fingerprints or codes are organized.

Filters are especially useful for RDMA hash tables. Because one remote RDMA access costs far more than one local memory access, a client can cache the filter locally and check it before touching the remote hash table. If the filter reports absence, the client ends the lookup immediately and saves one remote RDMA access. If the filter reports possible presence, the client still queries the remote hash table for confirmation. False positives let a few nonexistent keys still trigger remote accesses, but compared with always accessing the remote hash table, a filter sharply reduces the remote round trips spent on unnecessary lookups. For the capacity adjustment of filters and caches under dynamic workloads, the Bamboo filter improves the adaptivity and stability of filter resizing through smooth reconstruction~\cite{wang2024bamboo,wang2022bamboo}, and a lightweight working set size estimation method shows, from the angle of online cache capacity optimization, the importance of adjusting data structure capacity according to the access workload~\cite{gu2023adaptive}.

Table~\ref{tab:rdma-hash-comparison} further summarizes the main design features of representative RDMA-oriented hash table strategies. The comparison shows that recent designs increasingly move update operations to the client side and reduce remote CPU involvement, while support for resizing, filtering, caching, and write optimization differs across systems.

\begin{table*}[t]
\centering
\caption{Feature comparison of representative RDMA-oriented hash table designs.}
\label{tab:rdma-hash-comparison}
\resizebox{\textwidth}{!}{
\begin{tabular}{lccccccccc}
\toprule
\textbf{Work} & 
\textbf{Remote CPU} & 
\textbf{Memory Disaggregation} & 
\textbf{Client-side Update} & 
\textbf{Lock-free Concurrency} & 
\textbf{Dynamic Resizing} & 
\textbf{Fingerprint} & 
\textbf{Filter} & 
\textbf{Client Cache} & 
\textbf{Write Optimization} \\
\midrule
Pilaf   & $\checkmark$ & $\times$      & $\times$      & $\times$      & $\times$      & $\times$      & $\times$      & $\times$      & $\times$ \\
FaRM    & $\checkmark$ & $\times$      & $\times$      & $\times$      & $\times$      & $\times$      & $\times$      & $\times$      & $\times$ \\
DrTM    & $\checkmark$ & $\times$      & $\times$      & $\times$      & $\times$      & $\times$      & $\times$      & $\times$      & $\times$ \\
RACE    & $\times$      & $\checkmark$ & $\checkmark$ & $\checkmark$ & $\checkmark$ & $\checkmark$ & $\times$      & $\checkmark$ & $\times$ \\
SepHash & $\times$      & $\checkmark$ & $\checkmark$ & $\checkmark$ & $\checkmark$ & $\checkmark$ & $\checkmark$ & $\checkmark$ & $\checkmark$ \\
Outback & $\times$      & $\checkmark$ & $\checkmark$ & $\checkmark$ & $\times$      & $\times$      & $\times$      & $\checkmark$ & $\times$ \\
FUSEE   & $\times$      & $\checkmark$ & $\checkmark$ & $\checkmark$ & $\checkmark$ & $\times$      & $\times$      & $\checkmark$ & $\times$ \\
\bottomrule
\end{tabular}
}
\end{table*}

\section{Challenges and Opportunities}
\label{sec:challenges}

This section analyzes the main problems in hash table design for RDMA. For each challenge, we first describe the problem, then show through existing work that the problem exists, and finally discuss possible solutions.

\subsection{Number of Remote Accesses and Round-Trip Overhead}

In local memory, one hash table lookup usually touches only a few buckets or slots, so the access cost is low. When the hash table moves to RDMA remote memory, those local memory accesses become remote network accesses. If one hash operation reads several buckets, it issues several RDMA READs, which increases the lookup latency. The problem is worst in chained hashing and other pointer-linked structures: the address of the next access becomes known only after the previous remote read returns, so the accesses are serial, and the round trips stack up directly on the critical path of the operation. Existing collision-handling schemes all suffer from this remote access amplification under RDMA. Chained hashing reads the nodes along the chain one by one and cannot issue the reads in parallel. Open addressing probes several consecutive slots. Cuckoo hashing can compute its candidate locations in advance, but a triggered eviction may still cause several remote reads. Hash table design for RDMA therefore cannot focus on time complexity alone; it must also account for the number of remote accesses one operation actually issues, and for whether those accesses can run in parallel.

Existing RDMA hash tables reduce the remote access overhead by limiting the number of candidate buckets or by batching request submission. These methods, however, still leave the serial access problem unsolved. Future research can explore more predictable access paths, so that the remote locations needed by a lookup or an update are determined as much as possible before the operation starts. It can also study load-aware bucket layouts and collision-handling mechanisms that keep probe sequences short even under high load. The goal is to shrink the data volume of each access, reduce the number of remote accesses, and turn unparallelizable accesses into accesses that are predictable, batchable, or parallel.

\subsection{Trade-off Between One-Sided and Two-Sided Operations}

RDMA offers two classes of access: one-sided operations and two-sided operations. With one-sided operations, the client reads and writes remote memory directly, making them suitable for remote nodes with limited computing power; however, the client must know the remote data layout and implement the lookup, insertion, deletion, and consistency-check logic by itself. Two-sided operations let the remote CPU take part in hash table operations, which simplifies the client design and eases complex collision handling and concurrency control; however, they consume remote computing resources, and with many clients the server becomes the bottleneck of the whole system. A hash table for RDMA therefore has to choose between the two according to specific conditions, such as the available remote computing power. Prior systems illustrate different choices in this design space. HERD shows that server-assisted RDMA messaging can efficiently support small key-value operations, while FaSST demonstrates that two-sided RDMA datagram RPCs can support scalable distributed transactions~\cite{kalia2014herd,kalia2016fasst}. These systems suggest that the appropriate RDMA primitive depends on whether remote CPU cycles, client-side complexity, or network round trips are the dominant bottleneck.

One-sided operations lower the remote CPU overhead and improve scalability. However, relying on them exclusively creates new problems: the clients must implement all complex update logic, coordination among multiple clients becomes harder, and some operations need several remote reads and writes to complete. We believe the opportunity for future research lies in more flexible hybrid access mechanisms. Simple lookups and routine updates can keep using one-sided operations to minimize remote involvement. Resizing, collision handling on hot buckets, bulk migration, and complex consistency maintenance can instead use lightweight remote assistance or offloading to data processing units (DPUs) and SmartNICs. This preserves the low CPU overhead of one-sided operations while avoiding pushing all the complex logic onto the clients, and it strikes a better balance among performance, scalability, and implementation complexity.

\subsection{Concurrent Access and Data Consistency}

In a remote RDMA hash table, multiple clients may run lookups and insertions at the same time, so data consistency under concurrent access must be guaranteed. Unlike the local case, the remote CPU on disaggregated memory usually does not participate, so the clients can hardly rely on remote threads for locking or complex coordination. If remote locks are built on RDMA CAS, every lock and unlock costs a remote round trip, and heavy contention further causes waiting and retries, which greatly increases latency. Moreover, one-sided reads may interleave with concurrent writes, so a client may read a bucket or key-value data that is in the middle of an update; concurrent insertions of the same key may also cause duplicate writes or version conflicts. How to guarantee data consistency is therefore a central problem in RDMA hash table design. Beyond concurrency, a remote hash table on disaggregated memory also has to consider failure recovery and fault tolerance. Aceso studies efficient fault tolerance for key-value stores on disaggregated memory~\cite{hu2024aceso}, showing that in real systems, remote data structures must also be co-designed with recovery protocols, metadata consistency, and failure handling. Replication introduces another source of remote-write overhead. Rowan studies the replication of persistent-memory key-value stores and provides an RDMA abstraction that aggregates concurrent remote writes before placing them sequentially in persistent memory~\cite{wang2023rowan}. This design illustrates that consistency and fault tolerance must be considered together with RDMA access granularity, write amplification, and the ordering of concurrent updates.

Existing studies reduce remote synchronization overhead to some extent through lock-free access, self-verifying fields such as checksums, version information, and atomic updates. These methods, however, still cannot handle all complex concurrency scenarios. Future research can explore lightweight consistency protocols for RDMA, for example based on version numbers, checksums, epochs, or leases, so that a client can decide at low cost whether the remote data is in a stable state. Research can also pursue contention reduction for hot buckets, using write spreading, batched submission, append-based updates, or multi-version management to reduce CAS contention. For concurrent access during resizing, clearer mechanisms are needed for directory cache invalidation, migration state marking, and read/write forwarding, so that correctness is preserved while reducing remote synchronization and retry costs.

\subsection{RNIC Address Translation and Remote Memory Layout Constraints}

Each remote-memory request requires the RNIC to perform a metadata lookup that checks access permissions and resolves the requested virtual location to its corresponding physical address. To lower this cost, the RNIC caches part of the address translation information, but the cache capacity is limited. When the hash table is large and spans many remote memory pages, the RNIC's address-translation cache misses more often, which increases the access latency. On RDMA remote memory, hash table performance therefore depends on the memory registration method, the page granularity, the address contiguity, and the access locality.

Existing work usually mitigates this problem with huge-page registration, contiguous memory layouts, and compact bucket designs. Huge pages reduce the number of pages for the same capacity and thus the number of address mappings. Contiguous layouts reduce random accesses across regions. Compact buckets let a client read a whole bucket in one access and compare multiple slots locally. These methods reduce the address translation and request submission overhead, but they mainly fit relatively static layouts and fixed access patterns.

Future work should pursue remote memory layouts that are friendlier to the RNIC. On the one hand, hot buckets, directory entries, and frequently used metadata can be placed together to raise the hit rate of the RNIC's address-translation cache. On the other hand, hot and cold data can be separated by access frequency to reduce the pressure that large-scale random accesses put on the RNIC cache. Overall, the address translation limits show that RDMA hash table design cannot focus on hash collisions alone; it must include the RNIC hardware resources and the remote memory layout in its optimization objectives.

\subsection{Remote Resizing and Elastic Scaling}

A hash table usually needs to be resized once its load factor rises, and on RDMA remote memory, resizing is more complex than in local memory. First, resizing moves data over the network; with full-table rehashing, it consumes a large amount of RDMA bandwidth and sharply increases the tail latency. Second, multiple clients may have cached old directories, bucket addresses, or capacity information; after a resize these caches may become stale and cause wrong accesses or extra retries. Finally, resizing usually runs concurrently with normal lookups, insertions, and updates, and the system must keep the access results correct while the migration is still in progress. Remote resizing is therefore not only a capacity adjustment problem, but also a problem of data migration, cache consistency, and concurrency control.

Existing work eases the cost of full-table rebuilding through local resizing, segment-by-segment migration, and client-side validation, but elastic scaling under high concurrency and dynamic workloads remains difficult. Recent work continues to treat resizing as a first-class design objective. Shard explicitly targets scalable and resize-optimized hashing on disaggregated memory, showing that remote resizing remains an active research direction~\cite{zha2025shard}. At the key-value-store level, DINOMO considers elasticity and lightweight online reconfiguration on disaggregated persistent memory~\cite{lee2022dinomo}. Its design shows that elastic scaling is not limited to resizing an individual hash table: the system must also coordinate data ownership, client-side caching, replication, and index management when compute or memory resources change. Future research can explore finer-grained incremental resizing, so that a resize affects only a few hot buckets or subtables while normal requests run in parallel with the migration. Clearer migration state marking and directory cache update mechanisms are also needed, so that a client can tell whether a data item is at the old location, at the new location, or in migration. For workloads with strong load variation, a remote hash table should further support load-aware scaling and reorganization of hot and cold data, limiting the bandwidth consumption while reducing the impact of resizing on throughput and tail latency.

\section{Conclusion}
\label{sec:conclusion}

In this paper, we have surveyed hash table design for RDMA-based remote memory. Starting from the large-scale storage and fast lookup demands created by ever-growing IoT data, we explained why it makes sense to deploy hash tables on RDMA remote memory when a single server falls short in memory capacity and scalability. On this basis, we reviewed hash table research for RDMA remote memory and discussed how filters reduce unnecessary lookups. We then distilled five key challenges: the number of remote accesses and round-trip overhead, the trade-off between one-sided and two-sided operations, concurrent access and data consistency, RNIC address translation and remote memory layout constraints, and remote resizing and elastic scaling. Future work can study dedicated hash table structures for RDMA remote memory and explore how to balance fewer remote round trips, lower RNIC address translation overhead, lock-free concurrent access, and localized resizing, thereby exploiting the high bandwidth and low latency of RDMA more fully.

\bibliographystyle{unsrtnat}
\bibliography{references}

\end{document}